\documentclass[twocolumn,tighten]{aastex61}

\usepackage{amsmath,amssymb}

\shorttitle{Very compact millimeter sizes for composite star-forming/AGN SMGs}
\shortauthors{Ikarashi et al.}

\begin{document}

\title{Very compact millimeter sizes for composite star-forming/AGN submillimeter galaxies}

\correspondingauthor{Soh~Ikarashi}

\author{Soh~Ikarashi}
\affil{Kapteyn Astronomical Institute, University of Groningen, P.O.\ Box 800, 9700 AV Groningen, Netherlands} 
\email{sikarash@astro.rug.nl}

\author{Karina\,I.~Caputi}
\affiliation{Kapteyn Astronomical Institute, University of Groningen, P.O.\ Box 800, 9700 AV Groningen, Netherlands} 

\author{Kouji~Ohta} 
\affiliation{Department of Astronomy, Kyoto University, Kitashirakawa-Oiwake-Cho, Sakyo-ku, Kyoto 606-8502, Japan}

\author{R.\,J.~Ivison}
\affiliation{Institute for Astronomy, University of Edinburgh, Royal Observatory, Blackford Hill, Edinburgh EH9 3HJ, UK} 
\affiliation{European Southern Observatory, Karl Schwarzschild Str.\ 2, D-85748 Garching, Germany}

\author{Claudia\,D.~P.~Lagos}  
\affiliation{International Centre for Radio Astronomy Research, University of Western Australia, 7 Fairway, Crawley 6009, Perth WA, Australia}

\author{Laura~Bisigello}
\affiliation{Kapteyn Astronomical Institute, University of Groningen, P.O.\ Box 800, 9700 AV Groningen, Netherlands} 
\affiliation{SRON Space Research of Netherlands, 9747 AD, Groningen, The Netherlands} 

\author{Bunyo~Hatsukade}
\affiliation{Institute of Astronomy, University of Tokyo, 2-21-1 Osawa, Mitaka, Tokyo 181-0015, Japan}

\author{Itziar~Aretxaga}
\affiliation{Instituto Nacional de Astrof\'{\i}sica, \'Optica y Electr\'onica (INAOE), Luis Enrique Erro 1, Sta. Ma. Tonantzintla, 72840 Puebla, Mexico}

\author{James\,S.~Dunlop}
\affiliation{Institute for Astronomy, University of Edinburgh, Royal Observatory, Blackford Hill, Edinburgh EH9 3HJ, UK}

\author{David\,H.~Hughes}
\affiliation{Instituto Nacional de Astrof\'{\i}sica, \'Optica y Electr\'onica (INAOE), Luis Enrique Erro 1, Sta. Ma. Tonantzintla, 72840 Puebla, Mexico}

\author{Daisuke~Iono}  
\affiliation{National Astronomical Observatory of Japan, Mitaka, Tokyo 181-8588, Japan}
\affiliation{SOKENDAI (The Graduate University for Advanced Studies), Shonan Village, Hayama, Kanagawa 240-0193, Japan}

\author{Takuma~Izumi}
\affiliation{Optical and Infrared Astronomy Division, National Astronomical Observatory of Japan, Mitaka, Tokyo 181-8588, Japan} 

\author{Nobunari~Kashikawa}
\affiliation{SOKENDAI (The Graduate University for Advanced Studies), Shonan Village, Hayama, Kanagawa 240-0193, Japan}
\affiliation{Optical and Infrared Astronomy Division, National Astronomical Observatory of Japan, Mitaka, Tokyo 181-8588, Japan}

\author{Yusei~Koyama}
\affiliation{National Astronomical Observatory of Japan, Mitaka, Tokyo 181-8588, Japan}
\affiliation{SOKENDAI (The Graduate University for Advanced Studies), Shonan Village, Hayama, Kanagawa 240-0193, Japan}
\affiliation{Subaru Telescope, 650 North A’ohoku Place, Hilo, HI 96720, USA}

\author{Ryohei~Kawabe}
\affiliation{National Astronomical Observatory of Japan, Mitaka, Tokyo 181-8588, Japan}
\affiliation{SOKENDAI (The Graduate University for Advanced Studies), Shonan Village, Hayama, Kanagawa 240-0193, Japan}

\author{Kotaro~Kohno} 
\affiliation{Institute of Astronomy, University of Tokyo, 2-21-1 Osawa, Mitaka, Tokyo 181-0015, Japan} 
\affiliation{Research Center for the Early Universe, School of Science, University of Tokyo, 7-3-1 Hongo, Bunkyo, Tokyo 113-0033, Japan}

\author{Kentaro~Motohara}
\affiliation{Institute of Astronomy, University of Tokyo, 2-21-1 Osawa, Mitaka, Tokyo 181-0015, Japan}

\author{Kouichiro~Nakanishi}
\affiliation{National Astronomical Observatory of Japan, Mitaka, Tokyo 181-8588, Japan}
\affiliation{SOKENDAI (The Graduate University for Advanced Studies), Shonan Village, Hayama, Kanagawa 240-0193, Japan}

\author{Yoichi~Tamura}
\affiliation{Division of Particle and Astrophysical Science, Graduate School of Science, Nagoya University, Furo-cho, Chikusa-ku, Nagoya 464-8602, Japan}

\author{Hideki~Umehata}
\affiliation{The Open University of Japan, 2-11 Wakaba, Mihama-ku, Chiba 261-8586, Japan}

\author{Grant\,W.~Wilson}
\affiliation{Department of Astronomy, University of Massachusetts, Amherst, MA 01003, USA}

\author{Kiyoto~Yabe}
\affiliation{Kavli Institute for the Physics and Mathematics of the Universe (WPI), The University of Tokyo, 5-1-5 Kashiwanoha, Kashiwa, Chiba
277-8583, Japan}

\author{Min\,S.~Yun}
\affiliation{Department of Astronomy, University of Massachusetts, Amherst, MA 01003, USA}

\begin{abstract}
We report the study of far-IR sizes of submillimeter galaxies (SMGs) in relation to their dust-obscured star formation rate (SFR) and active galactic nuclei (AGN) presence, determined using mid-IR photometry.  We determined the millimeter-wave ($\lambda_{\rm obs}=1100\,\mu$m) sizes of 69 ALMA-identified SMGs, selected with $\geq10$$\sigma$ confidence on ALMA images ($F_{\rm 1100 \mu m}=1.7$--7.4\,mJy).  We found that all the SMGs are located above an avoidance region in the millimeter size-flux plane, as expected by the Eddington limit for star formation.  In order to understand what drives the different millimeter-wave sizes in SMGs, we investigated the relation between millimeter-wave size and AGN fraction for 25 of our SMGs at $z=1$--3. We found that the SMGs for which the mid-IR emission is dominated by star formation or AGN have extended millimeter-sizes, with respective median  $R_{\rm c,e} = 1.6^{+0.34}_{-0.21}$ and 1.5$^{+0.93}_{-0.24}$\,kpc. Instead, the SMGs for which the mid-IR emission corresponds to star-forming/AGN composites have more compact millimeter-wave sizes, with median  $R_{\rm c,e}=1.0^{+0.20}_{-0.20}$\,kpc. The relation between millimeter-wave size and AGN fraction suggests that this size may be related to the evolutionary stage of the SMG.  The very compact sizes for composite star-forming/AGN systems could be explained by supermassive black holes growing rapidly during the SMG coalescing, star-formation phase. 
\end{abstract}


\section{Introduction}

The morphology and size of star-forming regions in submillimeter galaxies (SMGs) are important properties with which we can address the nature of their prodigious, dust-obscured star formation, and consequently the formation and evolution of the most massive galaxies.  The Atacama Large Millimeter/submillimeter Array (ALMA) is enabling astronomers to image high-redshift SMGs with angular resolutions of $\lesssim0''$.3. Some ALMA studies have reported  effective radii ($R_{\rm e}$) of $\sim0.3$--3\,kpc \citep[e.g.][]{ika15,sim15,hod16}.  These radii are small compared with what astronomers expected from studies of SMG sizes based on radio continuum and CO emission \citep[e.g.][]{tac06,big08,ivi11}. These new results represent a new milestone in our understanding of star formation in SMGs, suggesting that these galaxies plausibly evolve to compact quiescent galaxies \citep[e.g.][]{tof14,ika15,sim15}. 

 As a next step, it would be useful to test the hypothesis that SMGs are connected to the formation of the most massive galaxies, being triggered by major mergers, and then evolving into compact quiescent galaxies via quenching in a QSO phase \citep[e.g.][]{san88,hop08,tof14}.  The compact submillimeter sizes of SMGs, including recent reports of the existence of subkilopersec-scale starburst cores \citep{ion16,ika17,ote17}, suggests that  the intense star-formation activity might be quenched by active galactic nuclei (AGN), as observed in some luminous QSOs \citep[e.g.][]{mai12,car16}. The link between SMGs and QSOs is still unclear, though. However, previous X-ray \citep[e.g.][]{ale05,wan13} and mid-IR \citep[e.g.][]{ivi04,cop10,ser10} studies indicate that some SMGs do harbor AGN.

In this letter, we report a millimeter-wave size study of 69 ALMA-identified AzTEC SMGs. Firstly, we study the empirical relation between the ALMA continuum flux densities and the millimeter-wave sizes of SMGs.  Secondly, we investigate the relationship between millimeter-wave sizes and the presence of AGN in SMGs at $z=1$--3, as determined from mid-IR data. We adopt throughout a cosmology with $H_{\rm 0}=70$\,km\,s$^{-1}$\,Mpc$^{-1}$, $\Omega_{\rm M}=0.3$, and $\Omega_{\rm \Lambda}=0.7$.

\section{ALMA Observations and samples} 

The sample used in this paper comes from our ALMA 1100-$\mu$m continuum imaging survey of 144 bright AzTEC/ASTE sources with $F_{\rm 1100 \mu m,\,AzTEC}\geq 2.4$\,mJy in the Subaru/{\it XMM-Newton} Deep Field \citep[SXDF;][]{fur08}.  The SXDF survey was conducted in the ALMA Cycles 2 and 3 (2013.1.00781, 2015.1.00442.S: PI. Hatsukade; B.\,Hatsukade\,et\,al.\,2017, in preparation).

The ALMA observations in Cycle 2 were carried out with the array configurations C34-5 and C34-7, with 37--38 working 12-m antennas covering up to a $uv$ distance of $\sim 1500$\,k$\lambda$.  In Cycle 3, the observations were executed in array configuration C40-4, covering up to a $uv$ distance of $\sim 1000$\,k$\lambda$.  On-source integration times per source in each cycle were 0.6\,min.  The typical synthesized beam size for our ALMA continuum images is $\sim 0.''30 \times 0.''23$ ($\rm PA \sim 56^{\circ}$), after combining the Cycle 2 and 3 data.  The average r.m.s.\ noise level is 120\,$\mu$Jy\,beam$^{-1}$.  The images were generated with Briggs weighting, using a robust parameter of 0.3.

The ALMA continuum maps yielded 70 ALMA-identified AzTEC SMGs (hereafter ASXDF SMGs) with $S_{\rm peak}/N\geq 10$ detections, suitable for reliable ALMA millimeter-wave size measurements \citep[e.g.][]{ika15}.  We removed one lensed SMG \citep[ASXDF1100.001;][]{ika11}, leaving 69 SMGs. ALMA fluxes were re-measured in tapered ALMA images with a synthesized beam of $\sim0{''}.6$, which is larger than the measured mm-wave sizes of SMGs in this paper, using the IMFIT task in CASA. 

For 51 ASXDF SMGs, we obtained well-constrained photometric redshifts, with a median error $\delta z= 0.13\pm0.02$, based on the individual 1-$\sigma$ errors estimated by {\it Le Phare} \citep[e.g.][]{ilb06} in spectral energy distribution (SED) model fitting using the $B$, $V$, $Rc$, $i'$, $z'$, $J$, $H$, $Ks$, 3.6 and 4.5\,$\mu$m data (S.\,Ikarashi et al.\ 2017, in preparation).  The remaining SMGs lie outside the coverage of the optical/near-IR images, or have individual 1-$\sigma$ errors of $>1$.  Photometric and spectroscopic redshifts from the literature are listed in Table~\ref{tbl-1}.

\section{ALMA millimeter-wave source size measurements}

We measured millimeter-wave sizes as circularized effective radii ($R_{\rm c,e}$) for the 69 ASXDF SMGs with ALMA visibility data, in the same manner as \citet{ika15}.  We used $uv$-distance versus amplitude plots (hereafter $uv$-amp plots) for our measurements.  Although the ALMA data cover $uv$ distances up to $\sim 1500$\,k$\lambda$, we used only data at $\leq 500$\,k$\lambda$, which corresponds to a scale of $\sim0.''2$.  Adopting this cutoff for the longest $uv$ distance is the equivalent of smoothing with a larger size kernel in the image plane.  We aim to mitigate the effects of possible  clumpy structures in the size measurements and to measure $R_{\rm c,e}$ robustly.  For the sources detected with $\geq10\sigma$ in the ALMA Cycle-2 images alone, we measured their sizes using only Cycle-2 data, to avoid effects due to any systematic absolute flux calibration offsets between our Cycle 2 and 3 data \footnote{Comparisons of the fluxes of ASXDF sources in our Cycle-1, 2 and 3 data indicated that the fluxes in the Cycle-3 data are systematically $\sim$20\% smaller. Therefore, we corrected the primary flux calibration for this effect.}.  We measured sizes by fitting a Gaussian model to the observed data in the $uv$-amp plots.
When we measure the size, the other sources ($\geq5\sigma$) in each ALMA image were removed from the visibility data based on simple source properties derived by IMFIT task. 

In order to estimate possible systematics in the size measurements, we injected mock sources into ALMA noise visibility images, generated from the actual ALMA data as in \citet{ika15}. 
Briefly we injected a symmetric Gaussian component for a range of source sizes and flux densities that
cover the putative parameter range of our ASXDF sources with uniform probability. 
As tested in \citet{ika15}, our method can measure circularized effective radii correctly even if a source has an asymmetric Gaussian profile. 
We corrected our raw measured sizes based on the results of the simulations for the data used in this paper. As a result, we obtained ALMA millimeter-wave sizes of 0$''.08$--0$''.68$ (FWHM) for the 69 ASXDF SMGs. 
Note that ASXDF1100.009.1 has two distinct millimeter-wave components with a separation of $\sim$0$''$.6, sharing a host galaxy at $z_{\rm spec}=0.9$.  

\section{Relation between millimeter sizes and fluxes}

\begin{figure*}
\epsscale{1.0}
\plotone{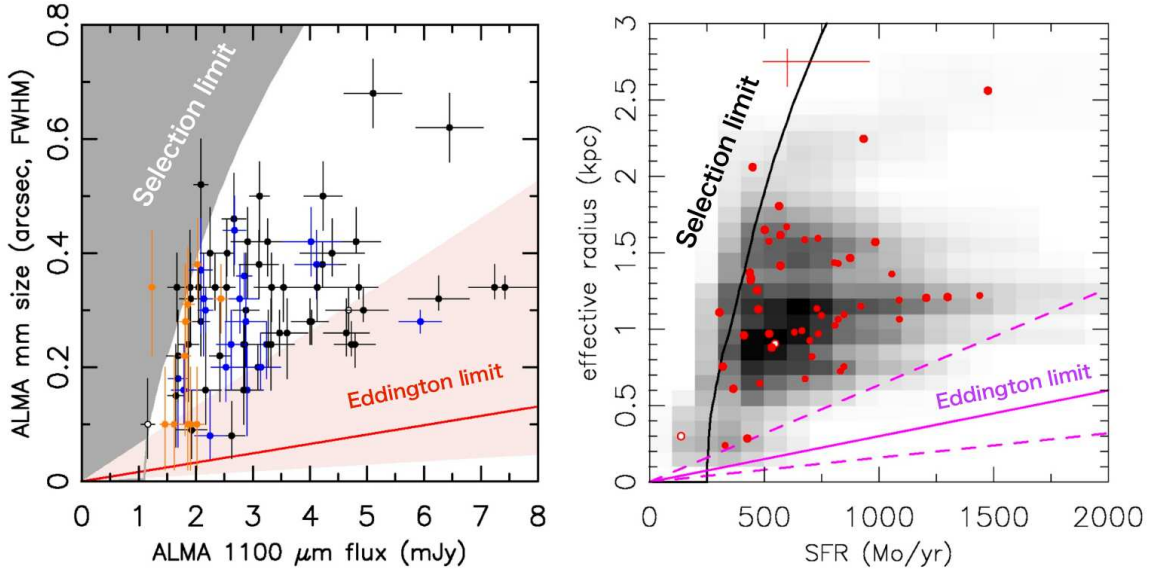}
\caption{
{\it Left:} ALMA 1100-$\mu$m flux vs.\ ALMA millimeter-wave size for the ASXDF SMGs with and without redshift information (filled black and blue   circles, respectively). 
The black points correspond to the ASXDF SMGs obtained in our ALMA Cycle-2 and 3 projects, as analyzed in this paper. 
The grey shaded area shows the approximate source selection limit (10$\sigma$) on our ALMA images. 
The orange points show other ASXDF SMGs at $z\gtrsim3$ from \citet{ika15}. 
The light red shaded area shows a range of Eddington-limited star formation for the 1$\sigma$ ranges of $T_d$ and gas velocity dispersion of known SMGs from the literature. The red solid line shows the Eddington-limited star-formation relation for the median $T_d$ and gas velocity dispersion.  
{\it Right:} SFR vs.\ effective radii in physical scale for the 51 ASXDF SMGs with available photometric or spectroscopic redshifts. 
The selection limit assumes a physical scale for $z=2$. 
The background grey-shaded area shows $P(SFR,size)$ for each SMG, taking into account the large uncertainty of the SFR due to the unknown dust temperature $T_d$. The Eddington-limit relation is indicated with magenta lines (solid for the median and dashed for $\pm1\sigma$ of the gas velosity dispersion).
Typical error bars are indicated with a red cross in the upper part of the plot.  
Open circles in both panels mark ASXDF1100.009.1, which has two distinct components in the ALMA image. 
}
\vspace{8mm}
\label{fig:sizeflux}
\end{figure*}

Fig.~\ref{fig:sizeflux} (left panel) shows all 69 ASXDF SMGs in an ALMA 1100-$\mu$m vs.\ millimeter-wave size plot.  Additionally, we plot 13 ALMA-identified, fainter SMGs at $z\gtrsim 3$ from \citet{ika15}.  ASXDF SMGs are absent from the top-left and the bottom-right corners of this plot.  The expected source selection limit for $\geq10\sigma$ continuum detection based on simple Gaussian models explains the absence of SMGs in the top-left corner.  The bottom-right corner, instead, is free from any such selection biases, so the absence of SMGs requires an explanation.

The absence of SMGs in the bottom-right corner of Fig. ~\ref{fig:sizeflux} can be interpreted as the influence of Eddington-limited star formation \citep{mur05}.  According to \citet{you08}, which reported pioneering studies of maximum star formation in bright SMGs, a maximum star-formation rate is given by \begin{equation} SFR_{\rm max} = 480\sigma^2_{400}D_{\rm kpc}\,\kappa^{-1}_{100} M_{\odot} yr^{-1}, \end{equation} where $D_{\rm kpc}$ is the characteristic physical scale of the starburst region in kpc, $\sigma_{\rm 400}$ is the line-of-sight gas velocity dispersion in units of 400\,km\,s$^{-1}$, and $\kappa_{\rm 100}$ is the dust opacity in units of 100\,cm$^2$\,g$^{-1}$.  Here we adopt a Chabrier initial mass function \citep{cha03}; $\kappa_{\rm 100}=1$, as in \citet{you08};  and a median gas velocity dispersion of 510\,km\,s$^{-1}$ from CO line observations of SCUBA SMGs \citep{bot13}. 
We also adopt 2$\times$ FWHM or 4$\times R_{\rm c,e}$, which is expected to include 94\% of the total light, as $D_{\rm kpc}$. The derived $SFR_{\rm max}$ was corrected with this factor of 0.94. 

In order to plot the relation between SFR and physical scale described by Equation~1 on Fig.~\ref{fig:sizeflux} (the left panel), we assume a fixed redshift $z=2$. The conversion factors from ALMA fluxes to SFRs were derived by bootstrapping given a dust temperature ($T_d$) distribution for lensed 1.3mm-selected galaxies \citep{wei13} and an SED library with $T_d$ information compiled in \citet{swi14}. 
For these assumptions, we obtain a possible range of Eddington-limited star formation rates.  

For a more direct comparison of the millimeter fluxes and sizes of SMGs with Eddington-limited star formation, we re-plot 51 of the 69 SMGs at $z=0.7$--6.8 with optical/near-IR photometric or spectroscopic redshifts on the SFR--physical size plane (Fig.~\ref{fig:sizeflux}, right panel).  
The SFRs are derived from $F_{\rm 1100 \mu m}$, given the range of possible dust temperatures $T_d$ and SEDs noted above. 
We assume that the AGN contribution to the submillimeter flux is negligible \citep[see references in][]{ros12}.  
In order to visualize the coverage of the size-SFR plane produced by the large SFR uncertainties (due to the unknown dust SED temperatures), we show 
the full SFR probability density distribution (rather than a single value) for each SMG. 
The results in both panels of Fig.~\ref{fig:sizeflux} show that the SMGs avoid the SFR region around the Eddington limit, suggesting that the minimum possible millimeter-wave sizes for bright SMGs are given by the Eddington limited star formation.

The empirical relation between flux and size can explain the apparent discrepancy between the reported (sub)millimeter-wave (median) sizes of $0.''20^{+0.''03}_{-0.''05}$ by \citet{ika15} and $0.''3\pm0.''04$ by \citet{sim15}.  Given $F_{\rm 870 \mu m}/F_{\rm 1100 \mu m}=2$ for conversion of the observed fluxes, \citet{sim15} covered $F_{\rm 1100 \mu m} \gtrsim 2.5$\,mJy. In this regime, our ASXDF SMGs have a median size of$0.''31^{+0.''03}_{-0.''03}$.

\section{Relation between millimeter sizes and AGN}\label{sec:agn}

\begin{figure}
\epsscale{1.0}
\plotone{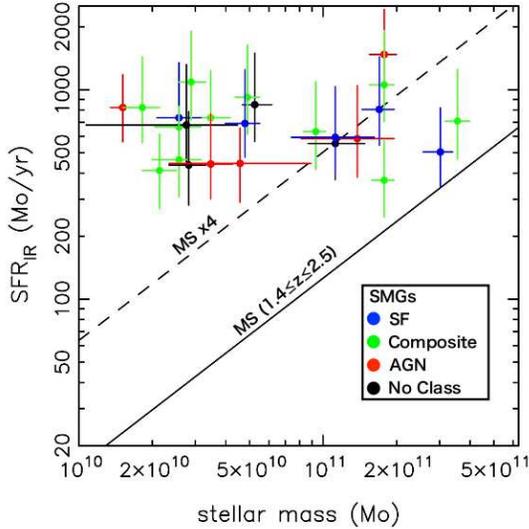}
\vspace{5mm}
\caption{Stellar mass vs.\ SFR for the 25 ASXDF SMGs at $z\sim$1--3. 
The black solid line corresponds to the the main sequence at $1.4\leq z\leq2.5$ \citep{dad07}. 
Colors indicate the AGN classification based on IRAC/MIPS colors (see \S5.1 and Fig.~\ref{fig:sizeagn} for details of this diagnostic).}
\label{fig:masssfr}
\end{figure}

We present our studies of the connection between the millimeter-wave sizes and AGN in SMGs, based on a mid-IR AGN diagnostic.  We consider 25 ALMA-identified SMGs with $1<z_{\rm phot\,or\,spec}<3$, which are detected in all IRAC and MIPS 24\,$\mu$m images.  
All SMGs here have redshift information and a single component at $\sim$0$''$.2 resolution. 
More than 15 out of the 25 are located above $4\times$ the main sequence at $z\sim2$ in the stellar mass vs.\ SFR plane (Fig.~\ref{fig:masssfr}), indicating that the majority of the sample are starbursts \citep{bis18}.  Note that among the 29 SMGs with $z=1$--3, four are not considered in our analysis: two SMGs are not detected at 24\,$\mu$m and the other two are blended in the IRAC maps.

\subsection{Mid-IR AGN diagnostic}

\begin{figure*}
\epsscale{1.0}
\plotone{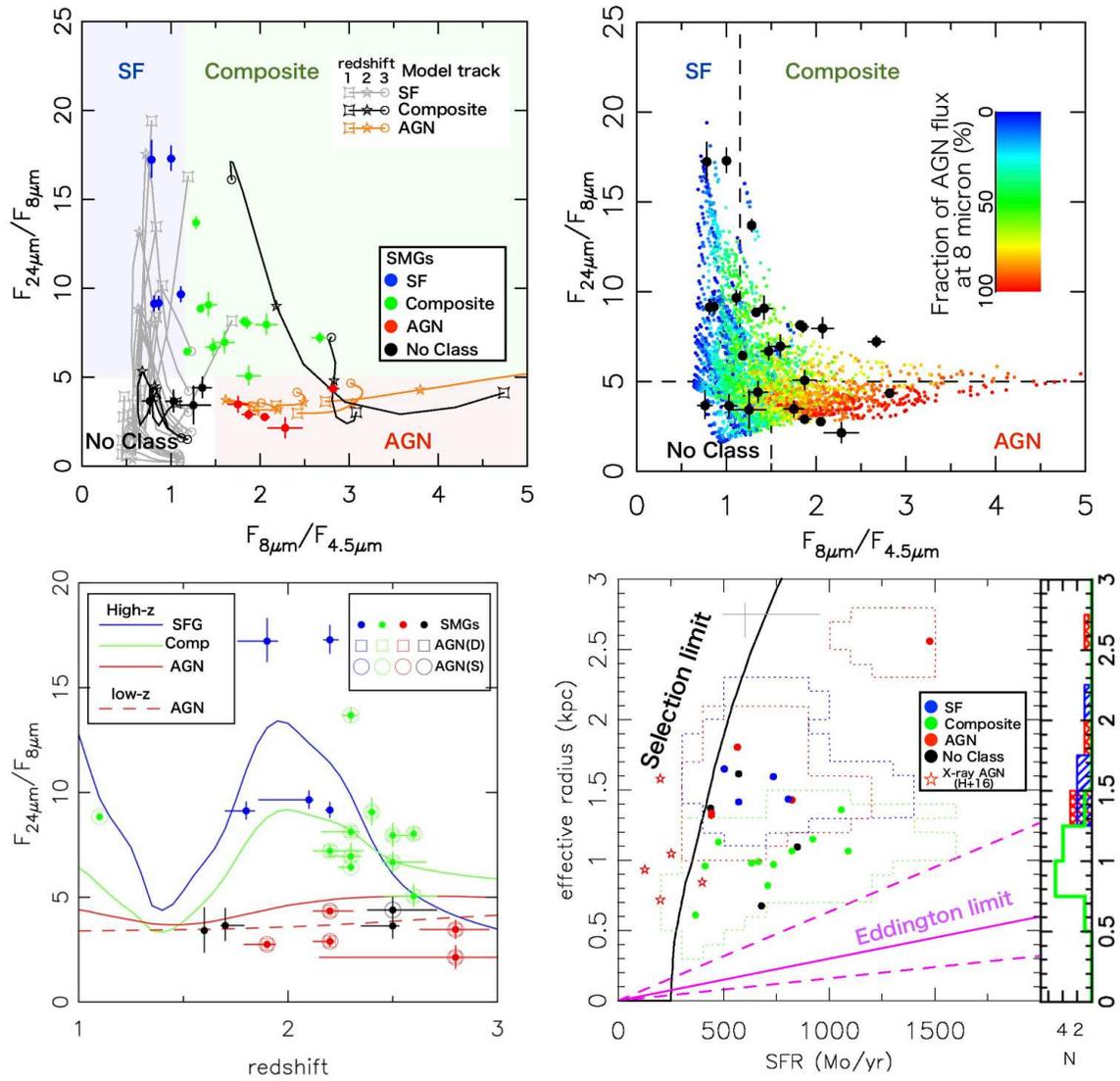}\vspace{3mm}
\caption{Relations between ALMA millimeter-wave size, SFR, and mid-IR color.  
{\it Top:} IRAC 4.5, 8, and MIPS 24-$\mu$m-color AGN diagnostic for $z\sim1$--3 galaxies, based on \citet{ivi04}. 
{\it Top left:} The colored shaded areas mark the diagnostic criteria of star-formation (SF) dominant, composite and AGN-dominant in mid-IR light. 
The solid curves are the predictions based on the SEDs in the SWIRE Template Library \citep{pol07}, which is mainly composed of local star-forming galaxies, (U)LIRGs, Seyfert galaxies, and QSOs. The colored filled circles indicate the ALMA-identified SMGs. 
{\it Top right:}
Simulated mid-IR colors of mock galaxies based on empirical SED templates, with the color points showing the AGN fraction based on the mock 8\,$\mu$m fluxes. 
The black points correspond to the ASXDF SMGs. 
The dashed lines show the criteria for SF/AGN classification. 
{\it Bottom left:}  Redshift versus 24-$\mu$m/8-$\mu$m colors for our sample. 
The solid lines indicate color evolution predictions based on empirical SED templates derived from star-forming-dominant, composite and AGN-dominant high-$z$ galaxy templates from \citet{kir15}. 
The dashed red line corresponds to an AGN-dominated system based on a local QSO SED template in the SWIRE template library. 
Open squares and circles indicate SMGs that satisfy the Donley et al. and Stern et al. IRAC AGN criteria, respectively. 
{\it Bottom right:} SFR vs. millimeter-wave effective radius.
The colored dotted lines delimit areas where $P(size,SFR)>=0.1\times P_{\rm peak}$ for each SF/AGN type. 
Host galaxies of X-ray-selected AGN from \citet[][]{har16} (H+16) are marked by red stars. The size distribution of our SMGs is shown in the histogram on the right-hand side of the plot. 
}
\label{fig:sizeagn}
\end{figure*}

A 4.5\,$\mu$m/8\,$\mu$m/24\,$\mu$m color-color plot has often been used as an AGN diagnostic for high-redshift, dusty infrared-luminous galaxies, such as SMGs and DOGs at $z\sim2$ \citep[e.g.][]{ivi04,ivi07,pop08a,pop08b}.  
We refer the reader to \citet{kir15}, who presented a detailed study of mid-IR SED evolution versus AGN fraction for high-$z$ galaxies. 
Empirical SED templates (top left panel in Fig.~\ref{fig:sizeagn}) suggest that high-redshift galaxies dominated by star formation or AGN in mid-IR light can be segregated from each other in the mid-IR color-color plane.  The position of our 25 SMGs in this color-color plot shows that some of them do not follow either the model tracks for star-formation-dominated or AGN-dominated galaxies.

We generated the expected mid-IR colors of galaxies that are a composite of SF and AGN by combining SEDs of SF and AGN with various SF/AGN ratios.  This `toy' color prediction reproduces the colors of `composite SMGs' which are likely to be dominated by neither an AGN nor a starburst in the mid-IR (top right panel in Fig.~\ref{fig:sizeagn}).

We divided the 25 SMGs into four sub-groups based on their 4.5/8/24-$\mu$m colors: star-forming, composite, AGN-dominant and `no class'. 
The criteria are:
\begin{itemize}
\item $F_{\rm 8 \mu m}/F_{\rm 4.5 \mu m}<1.15$ $\bigwedge$ $F_{\rm 24 \mu m}/F_{\rm 8 \mu m}\geq 5$ (star-forming)
\item $F_{\rm 8 \mu m}/F_{\rm 4.5 \mu m}\geq1.15$ $\bigwedge$ $F_{\rm 24 \mu m}/F_{\rm 8 \mu m}\geq 5$ (composite)
\item $F_{\rm 8 \mu m}/F_{\rm 4.5 \mu m}\geq1.50$ $\bigwedge$ $F_{\rm 24 \mu m}/F_{\rm 8 \mu m}< 5$ (AGN)
\item $F_{\rm 8 \mu m}/F_{\rm 4.5 \mu m}<1.50$ $\bigwedge$ $F_{\rm 24 \mu m}/F_{\rm 8 \mu m}< 5$ (no class).
\end{itemize}

The model colors (top, Fig.~\ref{fig:sizeagn}) indicate that the SMGs categorized as `no class' could be in the star-forming or composite classes.  Due to this ambiguity, we consider the `no class' separately.

Note that, In our diagnostic, the star-forming class and AGN dominant class are defined first. 
We choose  $F_{\rm 8.0 \mu m}/F_{\rm 4.5 \mu m}=$1.15 as criterion for separation, as this ensures that all galaxies that satisfy neither an AGN criteria by \citet{don12} nor another criteria by \citet{ste05} also lie on the star-forming region of the colour-colour diagram. 
The predicted 24$\mu$m/8$\mu$m color evolution with redshift, as derived by public empirical mid-IR SED templates for high-$z$ star-forming galaxies, composite galaxies, and AGN dominant galaxies \citep{kir15},  are shown along with our sample SMGs (bottom left, Fig.~\ref{fig:sizeagn}). For these templates, the respective mid-IR AGN fractions of each sample are $<$20, 20--80, and $\geq$80\%.  In this plot we averaged the public SEDs in each AGN class, after scaling all fluxes at $\lambda_{\rm rest}=8$\,$\mu$m. The predictions based on the Kirkpatrick et al. SED templates suggest that our criteria for 24$\mu$m/8$\mu$m color can work to select an AGN-dominant class, and show that our composite type is expected to have typically AGN fractions of around $\sim$50\%, consistently with our 'toy' models.

\subsection{Results}

In the millimeter-wave physical size vs.\ SFR plot (bottom right panel in Fig.~\ref{fig:sizeagn}), all SMGs with composite mid-IR components are evidently  more compact and located closer to the Eddington limit than the other SMGs with star-forming dominant or AGN dominant mid-IR components.

The respective median $R_{\rm c,e}$ for the SMGs classified as star-forming dominant, composites, and AGN dominant are 1.6$^{+0.34}_{-0.21}$, 1.0$^{+0.20}_{-0.20}$, and 1.5$^{+0.93}_{-0.24}$\,kpc. 
The size difference between the SMGs with composite and star-forming mid-IR components, and the difference between the SMGs with composite and AGN-dominant mid-IR components are real, with a significance level of $>99$\%, according to the Kolmogorov-Smirnov test. This indicates that the composite type galaxies are characterized by more compact star-forming regions than those of the star-forming or AGN-dominant types. 

We also explored the relation between size and stellar mass in our sample and found that the size differences are not a consequence of different stellar masses. Composite SMGs are the most compact of the three types, even at fixed stellar mass.

None of our ALMA-identified AzTEC SMGs are detected in the existing {\it XMM-Newton} X-ray maps \citep{ued08}, probably because these maps are too shallow. Nevertheless, we can compare our results with the sizes derived for the host galaxies of five high-$z$, X-ray-selected AGN ($L_{\rm 2-8keV}=10^{42.1-43.6}$\,erg\,s$^{-1}$) by \citet{har16}. These authors reported a size  distribution for their AGN hosts similar to the SMG sizes in \citet{sim15}.  The most X-ray luminous source in their sample (with $L_{\rm 2-8keV}=10^{43.6}$\,erg\,s$^{-1}$) has an extended size, and the remaining four ($L_{\rm 2-8keV}=10^{42.1-43.4}$\,erg\,s$^{-1}$) have compact sizes, which are comparable to those of our composite type here (Fig.~\ref{fig:sizeagn}, bottom right).

\subsection{AGN growth during a very compact star-forming phase?}

\begin{figure*}
\epsscale{1.0}
\plotone{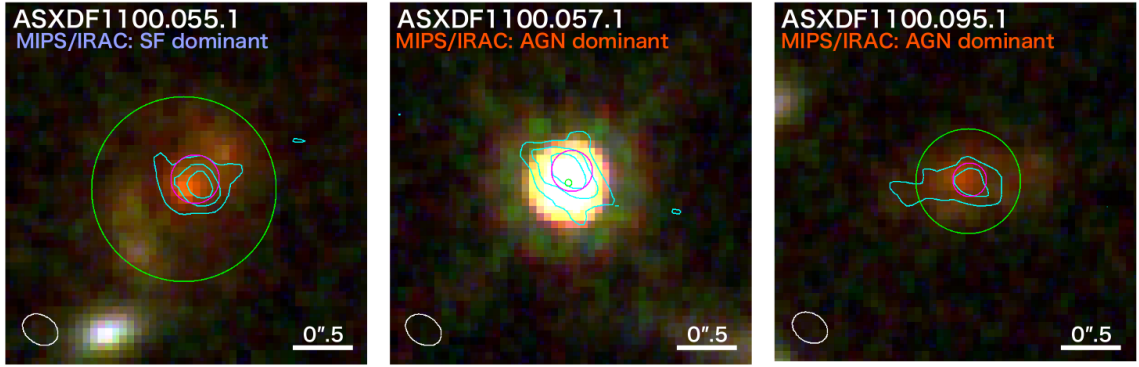}
\caption{{\it HST} images with ALMA continuum contours for three of our galaxies. 
All images are R(1.6$\mu$m)/G(1.2$\mu$m)/B(0.8$\mu$m) composites from CANDELS-UDS. 
The cyan contours correspond to the ALMA 1100-$\mu$m continuum (3, 5 and 7$\sigma$; 1$\sigma\sim$120$\mu$Jy\,beam$^{-1}$). 
The magenta circles indicate the circularized effective radii of ALMA 1100-$\mu$m continuum emission. 
The green circles correspond to the 1.6-$\mu$m continuum. The respective effective radii ($R_{\rm nir ,e}$) are 6.3$\pm$0.23, (240$\pm$1.7)$\times10^{-3}$, and 3.6$\pm$0.16\,kpc, for ASXDF1100.055.1, 057.1, and 095.1, after PSF deconvolution.    
The {\it HST} sizes are based on \citet{vdwel12}. }
\label{fig:hst}
\end{figure*}

The very compact millimeter-wave sizes of the SMGs with composite mid-IR components suggest that a central supermassive black hole could be growing in a compact and coalescing star-forming phase, which is consistent with the predictions of \citet{spr05} for galaxy major mergers. The extended millimeter-wave sizes of the SMGs of the star-forming dominant class can be explained by a mid-stage merger as seen in, e.g., VV114  \citep{sai15}.  Actually ASXDF1100.055.1 with the star-forming dominant class shows merger-like near-IR morphology (Fig.\,\ref{fig:hst}).  
Instead, the extended sizes of the  SMGs with the AGN-dominant class are puzzling. 
In line with the evolutionary scenarios of, e.g., \citet{san88,hop08,tof14} that SMGs evolve into QSOs, 
these extended sizes may be explained by positive AGN feedback by a growing supermassive black hole in the phase of star-formation quenching, as it is suggested by simulations for luminous AGN/QSOs \citep[e.g.][]{ish12,zub13} and considered for some luminous QSOs \citep[e.g.][]{car16}. 
In fact, ASXDF1100.057.1 with the AGN dominant class has a QSO-like near-IR morphology (Fig.\,\ref{fig:hst}). 
However, no significant near-IR morphological difference between AGN-host and non-AGN-host galaxies, that are not submillimeter selected, is reported \citep[e.g.][]{koc12}.  The extended submillimeter sizes in our SMGs may come from the nature of their host galaxies.

\begin{table*}
\begin{center}
\caption{Summary data of the ASXDF SMG sample analyzed in this paper.\label{tbl-1}}
\fontsize{5pt}{0pt}\selectfont
\begin{tabular}{ l c c c c c c c c c }
\tableline \tableline 
 ID                          &    R.A.           &     Dec.      &SNR& $F_{\rm 1100 \mu m}$  & $z_{\rm photo}$                         & SFR                               & mm-wave size      & mm-wave size       & AGN   \\  
                              &  (J2000)       &   (J2000)  &    &  (mJy)                                &                                                  & ($M_{\odot}$\,yr$^{-1}$)              & (FWHM; arcsec)          &  ($R_{\rm c,e}$; kpc)                                                             & (mid-IR)  \\  
 \tableline 
ASXDF1100.002.1 & 2:17:30.63 & -4:59:36.8 & 15 & 4.81$\pm$0.43 & 3.3$^{+0.07}_{-0.87}$ & 990$^{+720}_{-340}$ & 0.42$^{+0.06}_{-0.02}$ &  1.6$^{+0.2}_{-0.1}$ & \nodata  \\ 
ASXDF1100.004.1 & 2:18:05.65 & -5:10:49.7 & 14 & 4.39$\pm$0.56 & 3.5$^{+0.35}_{-0.16}$ & 880$^{+420}_{-290}$ & 0.40$^{+0.06}_{-0.04}$ &  1.5$^{+0.2}_{-0.1}$ & \nodata  \\ 
ASXDF1100.005.1 & 2:17:30.45 & -5:19:22.5 & 25 & 7.24$\pm$0.45 & 0.7$^{+0.01}_{-0.01}$ & 1200$^{+990}_{-420}$ & 0.34$^{+0.04}_{-0.02}$ &  1.2$^{+0.1}_{-0.1}$ & \nodata  \\ 
ASXDF1100.006.1 & 2:17:27.32 & -5:06:42.8 & 10 & 5.11$\pm$0.50 & 4.5$^{+0.18}_{-0.15}$ & 930$^{+340}_{-330}$ & 0.68$^{+0.06}_{-0.06}$ &  2.2$^{+0.2}_{-0.2}$ & \nodata  \\ 
ASXDF1100.007.1 & 2:18:03.01 & -5:28:42.0 & 20 & 6.26$\pm$0.53 & 3.2$^{+0.28}_{-0.22}$ & 1300$^{+930}_{-450}$ & 0.32$^{+0.04}_{-0.02}$ &  1.2$^{+0.1}_{-0.1}$ & \nodata  \\ 
ASXDF1100.008.1 & 2:16:47.93 & -5:01:29.9 & 12 & 6.45$\pm$0.59 & 2.2$^{+0.02}_{-0.08}$ & 1500$^{+950}_{-460}$ & 0.62$^{+0.06}_{-0.06}$ &  2.6$^{+0.2}_{-0.2}$ & AGN  \\ 
ASXDF1100.009.1A & 2:17:42.11 & -4:56:27.6 & 19 & 4.68$\pm$0.40 &(0.5)$^a$ & 550$^{+430}_{-190}$ & 0.30$^{+0.02}_{-0.04}$ &  0.9$^{+0.1}_{-0.1}$ & \nodata  \\ 
ASXDF1100.009.1B & 2:17:42.16 & -4:56:28.5 & 11 & 1.16$\pm$0.12 &(0.5)$^a$ & 140$^{+110}_{-50}$ & 0.10$^{+0.08}_{-0.06}$ &  0.6$^{+0.5}_{-0.4}$ & \nodata  \\ 
ASXDF1100.011.1 & 2:17:50.59 & -5:30:59.2 & 13 & 4.22$\pm$0.41 & 5.5$^{+0.08}_{-0.63}$ & 730$^{+440}_{-260}$ & 0.38$^{+0.04}_{-0.04}$ &  1.1$^{+0.1}_{-0.1}$ & \nodata  \\ 
ASXDF1100.014.1$^{\dagger}$  & 2:17:29.77 & -5:03:18.6 & 11 & 3.12$\pm$0.17 & 2.2$^{+0.04}_{-0.03}$ & 690$^{+270}_{-210}$ & 0.50$^{+0.06}_{-0.08}$ &  2.1$^{+0.2}_{-0.3}$ & SF  \\ 
ASXDF1100.016.1 & 2:16:41.11 & -5:03:51.4 & 19 & 4.79$\pm$0.35 & 5.0$^{+0.54}_{-0.06}$ & 850$^{+390}_{-240}$ & 0.24$^{+0.02}_{-0.04}$ &  0.8$^{+0.1}_{-0.1}$ & \nodata  \\ 
ASXDF1100.018.1 & 2:18:13.83 & -4:57:43.5 & 14 & 3.47$\pm$0.32 & 1.7$^{+0.09}_{-0.02}$ & 850$^{+650}_{-280}$ & 0.26$^{+0.04}_{-0.04}$ &  1.1$^{+0.2}_{-0.2}$ & NO  \\ 
ASXDF1100.020.1$^{\bullet}$ & 2:18:23.73 & -5:11:38.5 & 13 & 4.94$\pm$0.43 & 2.7$^{+0.01}_{-0.01}$ & 1100$^{+460}_{-380}$ & 0.30$^{+0.04}_{-0.02}$ &  1.2$^{+0.2}_{-0.1}$ & \nodata  \\ 
ASXDF1100.021.1 & 2:18:16.49 & -4:55:08.8 & 16 & 4.03$\pm$0.28 & 2.3$^{+0.03}_{-0.04}$ & 920$^{+720}_{-310}$ & 0.28$^{+0.02}_{-0.04}$ &  1.1$^{+0.1}_{-0.2}$ & COM  \\ 
ASXDF1100.022.1 & 2:18:42.68 & -4:59:32.1 & 15 & 3.09$\pm$0.31 & 2.3$^{+0.01}_{-0.06}$ & 710$^{+550}_{-240}$ & 0.20$^{+0.04}_{-0.04}$ &  0.8$^{+0.2}_{-0.2}$ & COM  \\ 
ASXDF1100.023.2 & 2:18:20.40 & -5:31:43.2 & 10 & 2.17$\pm$0.27 & 2.5$^{+0.10}_{-0.12}$ & 480$^{+350}_{-160}$ & 0.16$^{+0.10}_{-0.06}$ &  0.6$^{+0.4}_{-0.2}$ & \nodata  \\
ASXDF1100.025.2$^{\dagger}$ & 2:17:32.59 & -4:50:26.4 & 13 & 2.34$\pm$0.12 & 3.4$^{+0.16}_{-0.07}$ & 470$^{+320}_{-150}$ & 0.34$^{+0.06}_{-0.04}$ &  1.3$^{+0.2}_{-0.1}$ & \nodata  \\ 
ASXDF1100.029.1$^{\dagger}$ & 2:17:20.80 & -4:49:49.5 & 11 & 2.67$\pm$0.21 & 2.8$^{+0.16}_{-0.17}$ & 570$^{+360}_{-180}$ & 0.46$^{+0.08}_{-0.10}$ &  1.8$^{+0.3}_{-0.4}$ & AGN  \\ 
ASXDF1100.031.1$^{\dagger}$ & 2:17:37.24 & -4:47:53.0 & 13 & 2.09$\pm$0.15 & 2.5$^{+0.18}_{-0.12}$ & 480$^{+380}_{-170}$ & 0.28$^{+0.04}_{-0.06}$ &  1.1$^{+0.2}_{-0.2}$ & COM  \\ 
ASXDF1100.033.1 & 2:18:03.56 & -4:55:27.3 & 15 & 4.86$\pm$0.33 &  (2.6)$^c$ & 1100$^{+860}_{-350}$ & 0.34$^{+0.04}_{-0.02}$ &  1.4$^{+0.2}_{-0.1}$ & COM  \\ 
ASXDF1100.034.1 & 2:17:59.32 & -5:05:04.6 & 11 & 2.84$\pm$0.32 & (1.6)$^b$& 680$^{+640}_{-220}$ & 0.16$^{+0.08}_{-0.06}$ &  0.7$^{+0.3}_{-0.3}$ & \nodata  \\ 
ASXDF1100.035.1$^{\dagger,\bullet}$ & 2:17:35.37 & -5:28:37.3 & 12 & 2.09$\pm$0.12 & 2.7$^{+0.07}_{-0.11}$ & 450$^{+360}_{-150}$ & 0.52$^{+0.08}_{-0.08}$ &  2.1$^{+0.3}_{-0.3}$ & \nodata  \\ 
ASXDF1100.041.1 & 2:17:53.87 & -5:26:35.7 & 10 & 2.91$\pm$0.29 & 0.8$^{+0.00}_{-0.00}$ & 520$^{+260}_{-180}$ & 0.42$^{+0.06}_{-0.10}$ &  1.6$^{+0.2}_{-0.4}$ & \nodata  \\ 
ASXDF1100.042.1 & 2:18:38.29 & -5:03:18.3 & 12 & 3.26$\pm$0.40 & 3.2$^{+0.02}_{-0.01}$ & 680$^{+440}_{-240}$ & 0.42$^{+0.04}_{-0.06}$ &  1.6$^{+0.1}_{-0.2}$ & \nodata  \\ 
ASXDF1100.044.1 & 2:17:45.85 & -5:00:56.7 & 12 & 1.93$\pm$0.26 & 6.8$^{+0.20}_{-0.72}$ & 330$^{+210}_{-84}$ & 0.09$^{+0.07}_{-0.05}$ &  0.2$^{+0.2}_{-0.1}$ & \nodata  \\ 
ASXDF1100.046.1 & 2:17:13.34 & -4:58:57.4 & 16 & 4.00$\pm$0.32 & 3.5$^{+0.01}_{-0.10}$ & 810$^{+620}_{-280}$ & 0.28$^{+0.04}_{-0.04}$ &  1.0$^{+0.1}_{-0.1}$ & \nodata  \\ 
ASXDF1100.047.1$^{\dagger}$  & 2:17:56.73 & -4:52:39.0 & 11 & 2.25$\pm$0.17 & 2.2$^{+0.01}_{-0.02}$ & 500$^{+400}_{-160}$ & 0.40$^{+0.08}_{-0.06}$ &  1.6$^{+0.3}_{-0.2}$ & SF  \\ 
ASXDF1100.048.1$^{\dagger}$ & 2:17:46.16 & -4:47:47.2 & 14 & 2.55$\pm$0.11 & 2.5$^{+0.21}_{-0.12}$ & 570$^{+460}_{-200}$ & 0.40$^{+0.06}_{-0.04}$ &  1.6$^{+0.2}_{-0.2}$ & NO  \\ 
ASXDF1100.050.1$^{\star}$ & 2:18:22.30 & -5:07:37.0 & 11 & 3.32$\pm$0.40 & 3.0$^{+0.15}_{-0.15}$ & 700$^{+360}_{-240}$ & 0.24$^{+0.08}_{-0.08}$ &  0.9$^{+0.3}_{-0.3}$ & \nodata  \\ 
ASXDF1100.051.1$^{\dagger}$ & 2:18:23.96 & -5:32:07.8 & 12 & 2.63$\pm$0.23 & 0.7$^{+0.00}_{-0.04}$ & 430$^{+270}_{-150}$ & 0.08$^{+0.06}_{-0.04}$ &  0.3$^{+0.2}_{-0.1}$ & \nodata  \\ 
ASXDF1100.051.2$^{\dagger}$ & 2:18:24.59 & -5:31:48.5 & 11 & 2.88$\pm$0.23 & 4.7$^{+0.24}_{-0.15}$ & 520$^{+270}_{-160}$ & 0.30$^{+0.10}_{-0.06}$ &  1.0$^{+0.3}_{-0.2}$ & \nodata  \\ 
ASXDF1100.052.1$^{\dagger}$ & 2:17:33.17 & -5:01:54.5 & 11 & 2.05$\pm$0.14 & 2.8$^{+0.25}_{-0.65}$ & 440$^{+340}_{-150}$ & 0.34$^{+0.04}_{-0.06}$ &  1.3$^{+0.2}_{-0.2}$ & AGN  \\ 
ASXDF1100.055.1$^{\dagger}$ & 2:17:20.03 & -5:13:05.8 & 13 & 2.54$\pm$0.15 & 2.1$^{+0.02}_{-0.24}$ & 570$^{+290}_{-180}$ & 0.34$^{+0.06}_{-0.06}$ &  1.4$^{+0.2}_{-0.2}$ & SF  \\ 
ASXDF1100.057.1 & 2:17:32.41 & -5:12:50.9 & 12 & 3.54$\pm$0.38 & 1.9$^{+0.04}_{-0.11}$ & 820$^{+360}_{-260}$ & 0.34$^{+0.04}_{-0.06}$ &  1.4$^{+0.2}_{-0.3}$ & AGN  \\ 
ASXDF1100.076.1 & 2:16:41.04 & -5:01:12.5 & 13 & 4.13$\pm$0.55 & 4.8$^{+0.13}_{-0.41}$ & 750$^{+550}_{-230}$ & 0.34$^{+0.04}_{-0.06}$ &  1.1$^{+0.1}_{-0.2}$ & \nodata  \\ 
ASXDF1100.077.1$^{\dagger}$ & 2:18:11.00 & -4:49:51.9 & 12 & 1.69$\pm$0.20 & 4.1$^{+0.02}_{-0.12}$ & 320$^{+190}_{-110}$ & 0.22$^{+0.08}_{-0.08}$ &  0.8$^{+0.3}_{-0.3}$ & \nodata  \\ 
ASXDF1100.089.1 & 2:18:10.64 & -5:34:53.6 & 21 & 4.73$\pm$0.30 & 5.4$^{+0.11}_{-0.09}$ & 830$^{+600}_{-200}$ & 0.24$^{+0.04}_{-0.02}$ &  0.7$^{+0.1}_{-0.1}$ & \nodata  \\ 
ASXDF1100.095.1$^{\dagger}$ & 2:17:12.97 & -5:14:12.2 & 10 & 1.91$\pm$0.19 & 2.2$^{+0.11}_{-0.08}$ & 440$^{+320}_{-150}$ & 0.32$^{+0.08}_{-0.08}$ &  1.3$^{+0.3}_{-0.3}$ & AGN  \\ 
ASXDF1100.100.1 & 2:17:53.25 & -4:49:51.5 & 13 & 2.84$\pm$0.29 & 2.2$^{+0.16}_{-0.08}$ & 670$^{+550}_{-210}$ & 0.24$^{+0.04}_{-0.04}$ &  1.0$^{+0.2}_{-0.2}$ & COM  \\ 
ASXDF1100.105.1 & 2:18:02.86 & -5:00:31.6 & 13 & 2.86$\pm$0.30 &  (1.1)$^b$ & 630$^{+460}_{-220}$ & 0.24$^{+0.06}_{-0.08}$ &  1.0$^{+0.2}_{-0.3}$ & COM  \\ 
ASXDF1100.107.1$^{\dagger}$ & 2:18:07.85 & -5:25:49.3 & 11 & 1.67$\pm$0.16 & 4.6$^{+0.18}_{-0.86}$ & 310$^{+190}_{-80}$ & 0.34$^{+0.06}_{-0.06}$ &  1.1$^{+0.2}_{-0.2}$ & \nodata  \\ 
ASXDF1100.115.1 & 2:16:59.42 & -5:10:55.8 & 12 & 4.23$\pm$0.33 & (0.6)$^a$  & 600$^{+500}_{-220}$ & 0.50$^{+0.06}_{-0.06}$ &  1.7$^{+0.2}_{-0.2}$ & \nodata  \\ 
ASXDF1100.134.1 & 2:17:54.80 & -5:23:23.8 & 15 & 3.27$\pm$0.27 & 2.5$^{+0.16}_{-0.05}$ & 740$^{+500}_{-260}$ & 0.24$^{+0.06}_{-0.04}$ &  1.0$^{+0.2}_{-0.2}$ & COM  \\ 
ASXDF1100.156.1 & 2:16:38.33 & -5:01:21.5 & 11 & 3.33$\pm$0.31 & 1.8$^{+0.04}_{-0.10}$ & 810$^{+630}_{-260}$ & 0.34$^{+0.06}_{-0.06}$ &  1.4$^{+0.3}_{-0.3}$ & SF  \\ 
ASXDF1100.188.1$^{\dagger,\star}$ & 2:16:41.94 & -5:07:04.3 & 10 & 2.42$\pm$0.18 & 2.6$^{+0.28}_{-0.20}$ & 530$^{+450}_{-180}$ & 0.22$^{+0.10}_{-0.08}$ &  0.9$^{+0.4}_{-0.3}$ & \nodata  \\ 
ASXDF1100.203.1$^{\dagger}$  & 2:18:23.15 & -5:27:02.0 & 11 & 1.90$\pm$0.12 & 2.5$^{+0.03}_{-0.15}$ & 440$^{+330}_{-150}$ & 0.34$^{+0.10}_{-0.10}$ &  1.4$^{+0.4}_{-0.4}$ & NO  \\ 
ASXDF1100.227.1 & 2:17:44.27 & -5:20:08.6 & 24 & 7.42$\pm$0.57 & 3.7$^{+0.35}_{-0.14}$ & 1400$^{+760}_{-510}$ & 0.34$^{+0.02}_{-0.02}$ &  1.2$^{+0.1}_{-0.1}$ & \nodata  \\ 
ASXDF1100.228.1 & 2:18:09.66 & -5:18:43.1 & 12 & 3.11$\pm$0.34 & 1.9$^{+0.05}_{-0.14}$ & 740$^{+610}_{-240}$ & 0.38$^{+0.06}_{-0.06}$ &  1.6$^{+0.3}_{-0.2}$ & SF  \\ 
ASXDF1100.229.1 & 2:18:18.84 & -4:50:29.9 & 11 & 3.60$\pm$0.36 & 2.3$^{+0.05}_{-0.11}$ & 820$^{+620}_{-270}$ & 0.26$^{+0.06}_{-0.08}$ &  1.1$^{+0.2}_{-0.3}$ & COM  \\ 
ASXDF1100.235.1 & 2:17:36.00 & -5:20:34.4 & 13 & 4.64$\pm$0.40 & 2.3$^{+0.04}_{-0.14}$ & 1100$^{+820}_{-370}$ & 0.26$^{+0.06}_{-0.04}$ &  1.1$^{+0.2}_{-0.2}$ & COM  \\ 
ASXDF1100.236.1$^{\dagger}$ & 2:17:21.54 & -5:19:07.7 & 11 & 1.65$\pm$0.14 & 2.4$^{+0.02}_{-0.02}$ & 370$^{+250}_{-120}$ & 0.15$^{+0.09}_{-0.09}$ &  0.6$^{+0.4}_{-0.4}$ & COM  \\ 
ASXDF1100.247.1$^{\dagger}$  & 2:16:33.85 & -5:02:42.7 & 11 & 1.87$\pm$0.18 & 2.6$^{+0.11}_{-0.14}$ & 410$^{+260}_{-140}$ & 0.24$^{+0.08}_{-0.10}$ &  1.0$^{+0.3}_{-0.4}$ & COM  \\ 
ASXDF1100.003.1$^{\dagger}$  & 2:16:44.48 & -5:02:21.6 &15 &2.85$\pm$0.13 &\nodata & \nodata &0.36$^{+0.04}_{-0.04}$ &\nodata & \nodata   \\
ASXDF1100.010.1 & 2:17:39.79 & -5:29:19.2 &24 &5.94$\pm$0.37 &\nodata & \nodata &0.28$^{+0.02}_{-0.02}$ &\nodata & \nodata \\
ASXDF1100.026.1$^{\dagger}$  & 2:17:42.55 & -5:29:00.3 &11 &1.69$\pm$0.17 &\nodata & \nodata & 0.18$^{+0.06}_{-0.12}$&\nodata & \nodata   \\
ASXDF1100.040.1 & 2:17:55.24 & -5:06:45.1 &15 &3.14$\pm$0.35 &\nodata & \nodata &0.20$^{+0.06}_{-0.04}$&\nodata & \nodata  \\
ASXDF1100.053.1 & 2:16:48.20 & -4:58:59.6 &10 &4.02$\pm$0.51 &\nodata & \nodata &0.42$^{+0.06}_{-0.06}$ &\nodata & \nodata   \\
ASXDF1100.054.1 & 2:17:15.41 & -4:57:55.6 &11 &4.12$\pm$0.38 &\nodata & \nodata &0.38$^{+0.06}_{-0.06}$ &\nodata & \nodata   \\
ASXDF1100.068.1 & 2:17:42.17 & -5:25:46.8 &12 &3.24$\pm$0.30 &\nodata & \nodata &0.24$^{+0.04}_{-0.06}$&\nodata & \nodata   \\
ASXDF1100.070.1$^{\dagger}$  & 2:18:46.15 & -5:04:12.5 &12 &2.17$\pm$0.13 &\nodata & \nodata &0.30$^{+0.04}_{-0.06}$&\nodata & \nodata   \\
ASXDF1100.074.1 & 2:18:33.31 & -4:58:07.0 &10 &2.77$\pm$0.33 &\nodata & \nodata &0.32$^{+0.06}_{-0.06}$ &\nodata & \nodata  \\
ASXDF1100.097.1 & 2:18:18.54 & -5:34:34.7 &11 &2.53$\pm$0.26 &\nodata & \nodata &0.20$^{+0.08}_{-0.06}$ & \nodata & \nodata   \\
ASXDF1100.097.2$^{\dagger}$  & 2:18:17.61 & -5:34:27.9 &10 &2.14$\pm$0.26 &\nodata & \nodata &0.32$^{+0.08}_{-0.10}$& \nodata & \nodata   \\
ASXDF1100.133.1 & 2:18:05.51 & -5:35:46.5 &11 &2.25$\pm$0.26 &\nodata & \nodata &0.08$^{+0.08}_{-0.04}$ & \nodata & \nodata  \\
ASXDF1100.161.1$^{\dagger}$  & 2:18:13.76 & -5:37:27.3 &12 &2.68$\pm$0.20 &\nodata & \nodata &0.44$^{+0.06}_{-0.06}$ & \nodata & \nodata  \\
ASXDF1100.168.1 & 2:18:04.37 & -5:34:03.5 &11 &1.79$\pm$0.21 &\nodata & \nodata &0.16$^{+0.08}_{-0.06}$& \nodata & \nodata   \\
ASXDF1100.213.1$^{\dagger}$  & 2:18:44.02 & -5:35:31.3 &12 &2.90$\pm$0.28 &\nodata & \nodata &0.16$^{+0.08}_{-0.08}$& \nodata & \nodata   \\
ASXDF1100.231.1 & 2:17:59.65 & -4:46:49.8 &12 &2.88$\pm$0.36 &\nodata & \nodata &0.28$^{+0.08}_{-0.08}$ &\nodata & \nodata   \\
ASXDF1100.243.1$^{\dagger}$  & 2:16:50.43 & -5:10:16.2 &10 &2.09$\pm$0.20 &\nodata & \nodata &0.37$^{+0.09}_{-0.11}$&\nodata & \nodata  \\
ASXDF1100.252.1 & 2:17:05.65 & -5:15:04.9 &12 &2.62$\pm$0.25 &\nodata & \nodata &0.24$^{+0.06}_{-0.08}$ &\nodata & \nodata \\
 \tableline 
 \multicolumn{10}{l}{{\bf Notes.} $^{\dagger}$ ALMA flux, SNR, and size measurements are conducted in the ALMA data after combining the Cycle 2 and 3 data.} \\
 \multicolumn{10}{l}{For sources without $^{\dagger}$, all ALMA measurements were done in the ALMA Cycle-2 data.} \\ 
  \multicolumn{10}{l}{$^{\star}$ The SMGs are not included in the analysis in \S\,5 because of non-detection in 24\,$\mu$m.} \\
    \multicolumn{10}{l}{$^{\bullet}$ The SMGs are not included in the analysis in \S\,5 because of source blending in the IRAC maps.} \\ 
 \multicolumn{10}{l}{See \S\,5.1 for the columns of AGN.} \\
\multicolumn{10}{l}{$^{a}$spectroscopic redshifts by cross-identification with the UDS-z survey catalog \citep[e.g.][]{bra13,mcl13}.} \\
\multicolumn{10}{l}{$^b$spectroscopic redshifts by cross-identification with the SCUBA SMGs \citep{ban11}.} \\
\multicolumn{10}{l}{$^c$spectroscopic redshifts by cross-identification with the SCUBA SMGs \citep{cop10}.} \\
\end{tabular} 
 \end{center}
\end{table*}

\vspace{5mm}
\facilities{ALMA,Spitzer,Subaru,UKIRT,HST(STIS)}


\clearpage
\begin{thebibliography}{}
\bibitem[Alexander et al.(2005)]{ale05} Alexander, D.~M., Bauer, F.~E., Chapman, S.~C., et al.\ 2005, \apj, 632, 736 
\bibitem[Banerji et al.(2011)]{ban11} Banerji, M., Chapman, S.~C., Smail, I., et al.\ 2011, \mnras, 418, 1071 
\bibitem[Biggs \& Ivison(2008)]{big08} Biggs, A.~D., \& Ivison, R.~J.\ 2008, \mnras, 385, 893 
\bibitem[Bisigello et al. (2018)]{bis18} Bisigello, L., Caputi, K. I., Grogin, N., Koekemoer, A., 2018, \aap submitted (arXiv: 1706.06154) 
\bibitem[Bothwell et al.(2013)]{bot13} Bothwell, M.~S., Smail, I., Chapman, S.~C., et al.\ 2013, \mnras, 429, 304
\bibitem[Bradshaw et al.(2013)]{bra13} Bradshaw, E.~J., Almaini, O., Hartley, W.~G., et al.\ 2013, \mnras, 433, 194  
\bibitem[Carniani et al.(2016)]{car16} Carniani, S., Marconi, A., Maiolino, R., et al.\ 2016, \aap, 591, A28 
\bibitem[Chabrier(2003)]{cha03} Chabrier, G.\ 2003, \pasp, 115, 763 
\bibitem[Coppin et al.(2010)]{cop10} Coppin, K., Pope, A., Men{\'e}ndez-Delmestre, K., et al.\ 2010, \apj, 713, 503
\bibitem[Daddi et al.(2007)]{dad07} Daddi, E., Dickinson, M., Morrison, G., et al.\ 2007, \apj, 670, 156 
\bibitem[Donley et al.(2012)]{don12} Donley, J.~L., Koekemoer, A.~M., Brusa, M., et al.\ 2012, \apj, 748, 142 
\bibitem[Furusawa et al.(2008)]{fur08} Furusawa, H., Kosugi, G., Akiyama, M., et al.\ 2008, \apjs, 176, 1 
\bibitem[Harrison et al.(2016)]{har16} Harrison, C.~M., Simpson, J.~M., Stanley, F., et al.\ 2016, \mnras, 457, L122 
\bibitem[Hodge et al.(2016)]{hod16} Hodge, J.~A., Swinbank, A.~M., Simpson, J.~M., et al.\ 2016, \apj, 833, 103  
\bibitem[Hopkins et al.(2008)]{hop08} Hopkins, P.~F., Hernquist, L., Cox, T.~J., \& Kere{\v s}, D.\ 2008, \apjs, 175, 356-389 
\bibitem[Ikarashi et al.(2011)]{ika11} Ikarashi, S., Kohno, K., Aguirre, J.~E., et al.\ 2011, \mnras, 415, 3081 
\bibitem[Ikarashi et al.(2015)]{ika15} Ikarashi, S., Ivison, R.~J., Caputi, K.~I., et al.\ 2015, \apj, 810, 133
\bibitem[Ikarashi et al.(2017)]{ika17} Ikarashi, S., Ivison, R.~J., Caputi, K.~I., et al.\ 2017, \apj, 835, 286 
\bibitem[Ilbert et al.(2006)]{ilb06} Ilbert, O., Arnouts, S., McCracken, H.~J., et al.\ 2006, \aap, 457, 841 
\bibitem[Ishibashi \& Fabian(2012)]{ish12} Ishibashi, W., \& Fabian, A.~C.\ 2012, \mnras, 427, 2998
\bibitem[Ivison et al.(2004)]{ivi04} Ivison, R.~J., Greve, T.~R., Serjeant, S., et al.\ 2004, \apjs, 154, 124 
\bibitem[Ivison et al.(2007)]{ivi07} Ivison, R.~J., Greve, T.~R., Dunlop, J.~S., et al.\ 2007, \mnras, 380, 199
\bibitem[Ivison et al.(2011)]{ivi11} Ivison, R.~J., Papadopoulos, P.~P., Smail, I., et al.\ 2011, \mnras, 412, 1913
\bibitem[Iono et al.(2016)]{ion16} Iono, D., Yun, M.~S., Aretxaga, I., et al.\ 2016, \apjl, 829, L10 
\bibitem[Kirkpatrick et al.(2015)]{kir15} Kirkpatrick, A., Pope, A., Sajina, A., et al.\ 2015, \apj, 814, 9 
\bibitem[Kocevski et al.(2012)]{koc12} Kocevski, D.~D., Faber, S.~M., Mozena, M., et al.\ 2012, \apj, 744, 148
\bibitem[McLure et al.(2013)]{mcl13} McLure, R.~J., Pearce, H.~J., Dunlop, J.~S., et al.\ 2013, \mnras, 428, 1088 
\bibitem[Maiolino et al.(2012)]{mai12} Maiolino, R., Gallerani, S., Neri, R., et al.\ 2012, \mnras, 425, L66 
\bibitem[Murray et al.(2005)]{mur05} Murray, N., Quataert, E., \& Thompson, T.~A.\ 2005, \apj, 618, 569 
\bibitem[Oteo et al.(2017)]{ote17} Oteo, I., Zwaan, M.~A., Ivison, R.~J., Smail, I., \& Biggs, A.~D.\ 2017, \apj, 837, 182
\bibitem[Pope et al.(2008a)]{pop08a} Pope, A., Chary, R.-R., Alexander, D.~M., et al.\ 2008a, \apj, 675, 1171-1193 
\bibitem[Pope et al.(2008b)]{pop08b} Pope, A., Bussmann, R.~S., Dey, A., et al.\ 2008b, \apj, 689, 127-133
\bibitem[Polletta et al.(2007)]{pol07} Polletta, M., Tajer, M., Maraschi, L., et al.\ 2007, \apj, 663, 81
\bibitem[Rosario et al.(2012)]{ros12} Rosario, D.~J., Santini, P., Lutz, D., et al.\ 2012, \aap, 545, A45 
\bibitem[Saito et al.(2015)]{sai15} Saito, T., Iono, D., Yun, M.~S., et al.\ 2015, \apj, 803, 60 
\bibitem[Sanders et al.(1988)]{san88} Sanders, D.~B., Soifer, B.~T., Elias, J.~H., et al.\ 1988, \apj, 325, 74 
\bibitem[Serjeant et al.(2010)]{ser10} Serjeant, S., Negrello, M., Pearson, C., et al.\ 2010, \aap, 514, A10 
\bibitem[Springel et al.(2005)]{spr05} Springel, V., Di Matteo, T., \& Hernquist, L.\ 2005, \mnras, 361, 776 
\bibitem[Simpson et al.(2015)]{sim15} Simpson, J.~M., Smail, I., Swinbank, A.~M., et al.\ 2015, \apj, 799, 81  
\bibitem[Stern et al.(2005)]{ste05} Stern, D., Eisenhardt, P., Gorjian, V., et al.\ 2005, \apj, 631, 163
\bibitem[Swinbank et al.(2014)]{swi14} Swinbank, A.~M., Simpson, J.~M., Smail, I., et al.\ 2014, \mnras, 438, 1267 
\bibitem[Tacconi et al.(2006)]{tac06} Tacconi, L.~J., Neri, R., Chapman, S.~C., et al.\ 2006, \apj, 640, 228  
\bibitem[Toft et al.(2014)]{tof14} Toft, S., Smol{\v c}i{\'c}, V., Magnelli, B., et al.\ 2014, \apj, 782, 68
\bibitem[Ueda et al.(2008)]{ued08} Ueda, Y., Watson, M.~G., Stewart, I.~M., et al.\ 2008, \apjs, 179, 124-141  
\bibitem[van der Wel et al.(2012)]{vdwel12} van der Wel, A., Bell, E.~F., H{\"a}ussler, B., et al.\ 2012, \apjs, 203, 24 
\bibitem[Wang et al.(2013)]{wan13} Wang, S.~X., Brandt, W.~N., Luo, B., et al.\ 2013, \apj, 778, 179 
\bibitem[Wei{\ss} et al.(2013)]{wei13} Wei{\ss}, A., De Breuck, C., Marrone, D.~P., et al.\ 2013, \apj, 767, 88
\bibitem[Younger et al.(2008)]{you08} Younger, J.~D., Fazio, G.~G., Wilner, D.~J., et al.\ 2008, \apj, 688, 59-66 
\bibitem[Zubovas et al.(2013)]{zub13} Zubovas, K., Nayakshin, S., King, A., \& Wilkinson, M.\ 2013, \mnras, 433, 3079 
\end{thebibliography}
\end{document}